# Structural phase transitions and superconductivity induced in antiperovskite phosphide CaPd$_3$P

Akira Iyo,[1,*] Hiroshi Fujihisa,[1] Yoshito Gotoh,[1] Shigeyuki Ishida,[1] Hiroki Ninomiya,[1] Yoshiyuki Yoshida,[1] Hiroshi Eisaki,[1] Hishiro T. Hirose,[2] Taichi Terashima,[2] Kenji Kawashima,[1,3]

[1]National Institute of Advanced Industrial Science and Technology (AIST), Tsukuba, Ibaraki 3058568, Japan

[2]National Institute for Materials Science, Tsukuba, Ibaraki 3050003, Japan

[3]IMRA Material R&D Co., Ltd., Kariya, Aichi 4480032, Japan

## Corresponding Author

Akira Iyo, E-mail: iyo-akira@aist.go.jp

**ABSTRACT:** In this study, we succeeded in synthesizing new antiperovskite phosphides $M$Pd$_3$P ($M$ = Ca, Sr, Ba) and discovered the appearance of a superconducting phase (0.17 ≦ $x$ ≦ 0.55) in a solid solution (Ca$_{1-x}$Sr$_x$)Pd$_3$P. Three perovskite-related crystal structures were identified in (Ca$_{1-x}$Sr$_x$)Pd$_3$P and a phase diagram was built on the basis of experimental results. The first phase transition from centrosymmetric (*Pnma*) to non-centrosymmetric orthorhombic (*Aba*2) occurred in CaPd$_3$P near room temperature. The phase transition temperature decreased as Ca$^{2+}$ was replaced with a larger-sized isovalent Sr$^{2+}$. Bulk superconductivity at a critical temperature ($T_c$) of approximately 3.5 K was observed in a range of $x$ = 0.17–0.55; this was associated with the centrosymmetric orthorhombic phase. Thereafter, a non-centrosymmetric tetragonal phase (*I*4$_1$*md*) remained stable for 0.6 ≦ $x$ ≦ 1.0, and superconductivity was significantly suppressed as samples with $x$ = 0.75 and 1.0 showed $T_c$ values as low as 0.32 K and 57 mK, respectively. For further substitution with a larger-sized isovalent Ba$^{2+}$, namely (Sr$_{1-y}$Ba$_y$)Pd$_3$P, the tetragonal phase continued throughout the composition range. BaPd$_3$P no longer showed superconductivity down to 20 mK. Since the inversion symmetry of structure and superconductivity can be precisely controlled in (Ca$_{1-x}$Sr$_x$)Pd$_3$P, this material may offer a unique opportunity to study the relationship between inversion symmetry and superconductivity.

## INTRODUCTION

Superconductivity has significant research interest due to its potential practical applications, in addition to the profound physics behind superconductors. The discovery of new superconductors can considerably affect material science as well as the research on superconductivity. Most recently, we synthesized a new compound Mg$_2$Rh$_3$P and found that superconductivity can be induced at 3.9 K by introducing a small amount of Mg deficiency.[1,2] Mg$_2$Rh$_3$P can be regarded as an antiperovskite structure due to its crystallographic characteristic. That is, there is a three-dimensional network of corner-sharing Rh$_6$P octahedrons, and two Mg$^{2+}$ ions occupy the space surrounded by the octahedrons. The antiperovskite-type superconductors have attracted much attention due to their unique features *e.g.* a heavy-fermion superconductor with broken space inversion symmetry CePt$_3$Si ($T_c$ = 0.75 K),[3] a possible topological superconductor Sr$_{3-\delta}$SnO ($T_c$ = ~5 K),[4] spin-triplet superconductor Li$_2$Pt$_3$B ($T_c$ = ~2.8 K),[5,6] Ni-based compounds MgNi$_3$C ($T_c$ = 8 K),[7] ZnNi$_3$N ($T_c$ = ~3 K)[8] and CdNi$_3$C ($T_c$ = ~3 K).[8,9] Regarding antiperovskite related phosphide, $A$Pt$_3$P ($T_c$ = 8.4, 6.6, and 1.5 K for $A$ = Sr, Ca, and La, respectively),[10] SrPt$_6$P$_2$ ($T_c$ = 0.6 K)[11] and SrPt$_{10}$P$_4$ ($T_c$ = 1.4 K)[12] are known superconductors other than Mg$_2$Rh$_3$P.

Synthesis of phosphides is relatively difficult due to the high vapor pressure and high reactivity of P at high temperatures. We speculated that this may have hindered the exploration of new compounds for antiperovskite phosphides. Our intensive search for new antiperovskite phosphides resulted in the successful synthesis of $M$Pd$_3$P ($M$ = Ca, Sr, Ba). SrPd$_3$P and BaPd$_3$P exhibited no superconductivity above 2 K. On the other hand, CaPd$_3$P exhibited slight superconductivity at ~4 K. Furthermore, CaPd$_3$P had a different crystal structure than that of the other two compounds, which motivated us to synthesize a solid solution between CaPd$_3$P and SrPd$_3$P. As a result, we detected structural phase transitions and bulk superconductivity in the solid solution (Ca$_{1-x}$Sr$_x$)Pd$_3$P. Herein, we present the phase diagram of (Ca$_{1-x}$Sr$_x$)Pd$_3$P, which was constructed based on experimental results, such as those of X-ray diffraction (XRD) analysis and resistivity and magnetization measurements, as functions of composition and temperature.

We first show X-ray diffraction (XRD) patterns and lattice parameters of the samples as functions of composition and temperature. Next, we indicate three types of crystal structures that appeared in (Ca$_{1-x}$Sr$_x$)Pd$_3$P. We also demonstrate how the structural phase transition temperature changed and how superconductivity developed in (Ca$_{1-x}$Sr$_x$)Pd$_3$P, indicating the composition dependence of resistivity and magnetization data.

Finally, we present the phase diagram of $(Ca_{1-x}Sr_x)Pd_3P$ constructed basis the experimental results.

## EXPERIMENTAL

### Material synthesis

Polycrystalline samples were synthesized by solid state reaction. A Pd powder (Kojundo Chemical, 99.9%) and precursors with nominal compositions of CaP, SrP, and BaP were used as starting materials. The precursors were obtained by reacting Ca (Furuuchi Chemical, 99.5%), Sr (Furuuchi Chemical, 99.9%), or Ba (Rare Metallic, 99%) with P (Furuuchi Chemical, 99.999%) at 700°C for 24 h in evacuated quarts tubes. The Ca, Sr, or Ba were processed into flaky shapes prior to the synthesis of precursors to accelerate the reaction. A sample with a nominal composition of $(Ca_{1-x}Sr_x)Pd_3P$ or $(Sr_{1-y}Ba_y)Pd_3P$ was ground using a mortar. The ground powder was pressed into a pellet (~0.15 g), which was then enclosed in an evacuated quartz tube (inner diameter of 8 mm, length of ~70 mm). The sample was heated at 970°C for 12 h followed by furnace cooling. Raw materials and precursors were processed in an $N_2$-filled glove box to prevent oxidation. Samples of $(Ca_{1-x}Sr_x)Pd_3P$ and $(Sr_{1-y}Ba_y)Pd_3P$ remained stable in air for at least several months.

### Measurements

Powder XRD patterns were obtained at room temperature (RT) (~293 K) using a diffractometer (Rigaku, Ultima IV) with Cu$K\alpha$ radiation. Powder XRD patterns of CaPd$_3$P were also measured at 150 K using a laboratory made cooling system. The obtained diffraction patterns were indexed and the space groups were searched using BIOVIA Materials Studio (MS) X-Cell software.[13] TOPAS version 5 software was used to search for the initial atomic coordinates.[14] Pawley and Rietveld analyses were performed using BIOVIA MS Reflex version 2020 software.[15] We removed the constraints of the space groups from the candidate models and obtained stable atomic coordinates using DFT calculations with MS Castep software,[16] while confirming that the original space groups had been preserved.

Magnetization ($M$) measurements were performed under a magnetic field ($H$) of 10 Oe, using a magnetic-property measurement system (Quantum Design, MPMS-XL7). Measurements were performed with zero-field-cooling (ZFC) and field-cooling (FC) modes. Electrical resistivity in the temperature range 2–320 K were measured by a four-probe method, using a physical property measurement system (Quantum Design, PPMS). Electrical resistivities of $(Sr_{0.75}Ca_{0.25})Pd_3P$, SrPd$_3$P, and BaPd$_3$P were measured down to 20 mK using a dilution refrigerator.

## RESULTS AND DISCUSSION

### Powder XRD patterns at RT

Three types of crystal structures appeared depending on the composition and temperature in $(Ca_{1-x}Sr_x)Pd_3P$ and $(Sr_{1-y}Ba_y)Pd_3P$. Figure 1 shows the powder XRD patterns at RT for representative compositions ($x$ = 0, 0.05, 0.25, 0.55, 0.6, 0.75, and 1.0, $y$ = 0.5 and 1.0). CaPd$_3$P ($x$ = 0) has a more complex diffraction pattern than that of others due to mixing of centrosymmetric (CS) and non-centrosymmetric (NCS) orthorhombic phases as described in subsection 4.3. Only a small amount of the Sr substitution ($x$ = 0.05) made the CS-orthorhombic phase dominant. The CS-orthorhombic phase was stable up to $x$ = 0.55. Thereafter, the XRD pattern clearly changed for $x$ = 0.6 due to a structural phase transition from CS-orthorhombic to NCS-tetragonal. The NCS-tetragonal phase was stable for a composition range $0.6 \leq x \leq 1$ and continued for a whole composition range of $(Sr_{1-y}Ba_y)Pd_3P$. Figure 2 shows the composition dependence of reduced lattice parameters and reduced cell volume ($V_r$) for the CS-orthorhombic ($0 \leq x < 0.6$) and NCS-tetragonal ($0.6 \leq x \leq 1.0$) phases in $(Ca_{1-x}Sr_x)Pd_3P$. The lattice parameters and $V_r$ increased almost linearly according to Vegard's law in each phase as the substitution of the larger size isovalent $Sr^{2+}$ for $Ca^{2+}$ progressed, which ensured that systematic samples were successfully prepared. Lattice parameters and $V_r$ of $(Sr_{1-y}Ba_y)Pd_3P$ (not shown) also increased as a proportion of larger $Ba^{2+}$ ions ($y$) increased. The $V_r$ increased by nearly 10.4 % from CaPd$_3$P ($V_r$ = 88.36 Å$^3$) to BaPd$_3$P ($V_r$ = 97.51 Å$^3$). The Pd$_6$P octahedron network appears to be flexible with respect to the size of the cations acceptable to form crystals. Therefore, other cations such as alkali metals and lanthanides may also be accepted into the Pd$_6$P network.

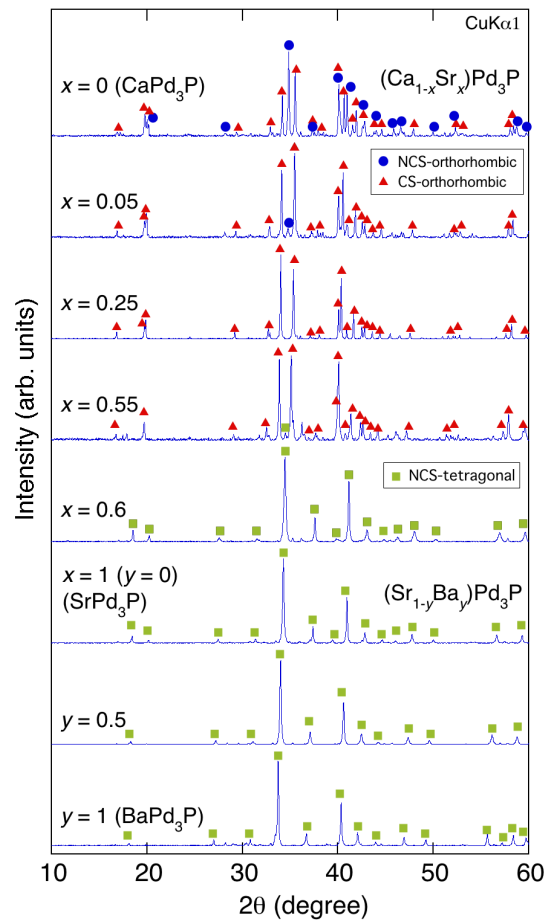

Figure 1. Powder XRD patterns at RT for representative compositions of $(Ca_{1-x}Sr_x)Pd_3P$ and $(Sr_{1-y}Ba_y)Pd_3P$. Major diffraction peaks from the NCS- and CS-orthorhombic and NCS-tetragonal phases are indicated by closed circles, triangles and squares, respectively. The contribution of Cu$K\alpha_2$ to the diffraction data was eliminated through software processing.

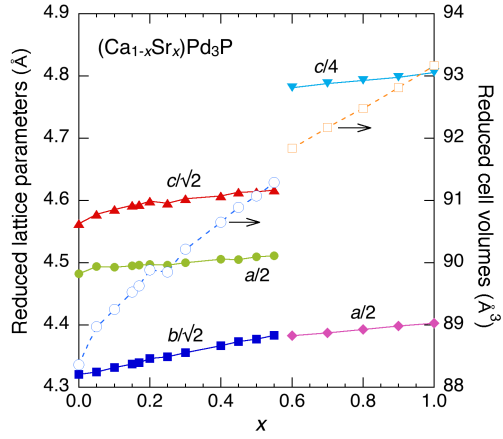

Figure 2. Composition $x$ dependence of reduced lattice parameters and reduced cell volumes for CS-orthorhombic ($0 \leq x < 0.6$) and NCS-tetragonal ($0.6 \leq x \leq 1.0$) phases in $(Ca_{1-x}Sr_x)Pd_3P$.

## Powder XRD pattern of CaPd$_3$P at low temperature

Figure 3 shows the powder XRD patterns of CaPd$_3$P at 150 K with those at RT. The diffraction pattern at RT could be analyzed assuming the existence of the two orthorhombic phases. The diffraction peaks from the CS-orthorhombic phase almost disappeared at 150 K. Note that 150 K is sufficiently lower than the temperatures (269 and 283 K) at which CaPd$_3$P exhibited hysteretic behavior in resistivity due to the structural phase transition as shown in the section 5. Note that a phase transition from a CS to an NCS structure with decreasing temperature (~200 K) is also observed in the pyrochlore superconductor Cd$_2$Re$_2$O$_7$ ($T_c$ = 1K).[17] In Cd$_2$Re$_2$O$_7$, an additional structural phase transition occurs at ~120 K. The structural changes are attributed to the distortion of Re tetrahedra.

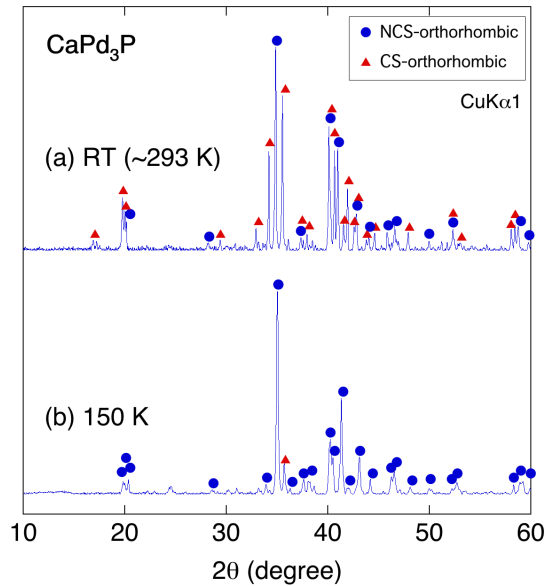

Figure 3. Powder XRD patterns of CaPd$_3$P at (a) RT and (b) 150 K. The contribution of CuK$\alpha_2$ to the diffraction data was eliminated through software processing.

## Crystal structure refinements

Crystal structures were refined for the representative samples of the NCS-orthorhombic (150 K) CaPd$_3$P, CS-orthorhombic (Ca$_{0.75}$Sr$_{0.25}$)Pd$_3$P, and NCS-tetragonal SrPd$_3$P (BaPd$_3$P). As shown in Figure 3 (a), CaPd$_3$P contained both NCS- and CS-orthorhombic phases, which complicated the structure refinement (only the lattice parameters could be determined). Therefore, the structure of CaPd$_3$P was refined at a low temperature (150 K). Rietveld fittings are shown for the three samples in Figure 4, and the refined structure parameters are summarized in Tables 1–3. For CaPd$_3$P at 150 K, we used the Pawley method to find that its crystal system is orthorhombic with a base-centered symmetry. The diffraction intensities could be fitted to the CS-orthorhombic space group $Cmce$ ($a$ = 8.75 Å, $b$ = 8.99 Å, $c$ = 8.93 Å) and the NCS-orthorhombic $Aba2$ ($a$ = 8.93 Å, $b$ = 8.99 Å, $c$ = 8.75 Å). $R_{wp}$ of the Rietveld refinement for the former model was 2.2% higher than that for the latter. An enthalpy comparison based on DFT calculations showed that the value for the former was 0.31 eV higher than that for the latter. For these reasons, we concluded that the NCS-orthorhombic $Aba2$ was the correct space group for CaPd$_3$P at 150 K. Note that $R_{wp}$ for CaPd$_3$P at 150 K (24%) was worse than those for the other samples because of the use of a sample cover to protect the CaPd$_3$P from frost. The cover decreased the peak intensity below diffraction angles of 30°. However, we believe that the atomic coordinates are reliable because the fitting between 30° and 140° in the Rietveld analysis is as satisfactory as in the other analyses. In the analysis of (Ca$_{0.75}$Sr$_{0.25}$)Pd$_3$P, $R_{wp}$ decreased by only 0.4% when using individual isotropic atomic displacement parameter $U_{iso}$ per atomic site. This resulted in $U_{iso}$ values of 0.009, 0.012, and 0.046 for cations, P, and Pd, respectively. The cation $U_{iso}$ was considered to be too small while that of the Pd was too large for 150 K. Therefore, we fixed $U_{iso}$ for all elements. Crystal structures are schematically drawn for CaPd$_3$P, (Ca$_{0.75}$Sr$_{0.25}$)Pd$_3$P and SrPd$_3$P in Figure 5. They commonly have three-dimensional networks of corner sharing Rh$_6$P octahedrons. However, as is often seen in perovskite structures, their space groups changed due to the deformation and rotation of the Rh$_6$P octahedrons.

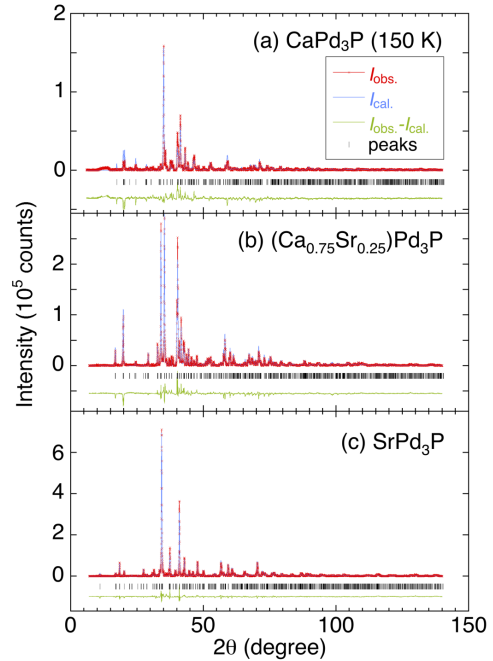

Figure 4. Rietveld fittings for (a) NCS-orthorhombic CaPd$_3$P (150 K), (b) CS-orthorhombic (Ca$_{0.75}$Sr$_{0.25}$)Pd$_3$P, and (c) NCS-tetragonal SrPd$_3$P. $I_{obs}$ and $I_{cal}$ indicate the observed and calculated diffraction intensities, respectively. The peaks from Cu$K\alpha_2$ line were not eliminated for the Rietveld analyses.

CaPd$_3$P had the lowest symmetric structure with a space group of $Aba2$ among them. Compared to a simple perovskite structure, periodicity is doubled along each three-axis direction (formula unit $Z = 8$). (Ca$_{0.75}$Sr$_{0.25}$)Pd$_3$P ($Pnma$) can be considered to be related to BaAu$_3$Ge(SrPt$_3$P)-type structure ($P4/nmm$). That is, P atoms adjacent in the $b$- and $c$-axes direction are alternately displaced in the opposite direction in the $b$-$c$ plane. Both $Pnma$ and $P4/nmm$ belong to a CS space group. The distortion of the Pd$_6$P octahedrons causes $\sqrt{2}$ and 2 times periodicity ($Z = 4$) along the $b(c)$-axis and the $a$-axis directions, respectively.

On the other hand, SrPd$_3$P and BaPd$_3$P ($I4_1md$) can be understood based on the NCS CePt$_3$B(CePt$_3$Si)-type structure ($P4mm$). That is, P atoms are displaced in the same direction along the $c$-axis. For this reason, the crystal structure of SrPd$_3$P and BaPd$_3$P lacks an inversion symmetry. Periodicity is doubled in the $a$- and $b$-axis directions and quadrupled in the $c$-axis direction ($Z = 16$).

**Table 1. Results of Rietveld structure refinement for CaPd$_3$P at 150 K.**

| atoms | Wyckoff position | x | y | z |
|---|---|---|---|---|
| Ca1 | 4a | 0 | 0 | 0.2496 |
| Ca2 | 4a | 0 | 0 | 0.7800(20) |
| Pd1 | 8b | 0.2742(6) | 0.0378(5) | 0.0299(16) |
| Pd2 | 8b | 0.2026(5) | 0.2461(7) | 0.2688(18) |
| Pd3 | 8b | 0.0328(5) | 0.2743(6) | -0.0027(17) |
| P | 8b | 0.2903(13) | 0.2795(13) | 0.0319(22) |

$Aba2$ (orthorhombic no. 41), $a = 8.9224(7)$ Å, $b = 8.9841(7)$ Å, $c = 8.7470(16)$ Å, $V = 701.2(1)$ Å$^3$ ($Z = 8$), $R_{wp} = 24.00\%$, $R_e = 11.74\%$, $S = 2.04$, $U_{iso} = 0.026(1)$ Å$^2$ for all atoms. The occupancy for each atom was fixed at 1.

**Table 2. Results of Rietveld structure refinement for (Ca$_{0.75}$Sr$_{0.25}$)Pd$_3$P at RT.**

| atoms | Wyckoff position | x | y | z |
|---|---|---|---|---|
| CaSr | 4c | -0.0016(7) | 1/4 | 0.7832(7) |
| Pd1 | 4c | 0.4568(3) | 1/4 | 0.2040(4) |
| Pd2 | 8d | 0.2268(2) | 0.0145(5) | 0.4554(2) |
| P | 4c | 0.1935(9) | 1/4 | 0.1693(11) |

$Pnma$ (orthorhombic no. 62), $a = 8.9944(3)$ Å, $b = 6.1545(2)$ Å, $c = 6.4989(2)$ Å, $V = 359.7(1)$ Å$^3$ ($Z = 4$), $R_{wp} = 12.37\%$, $R_e = 7.76\%$, $S = 1.59$, $U_{iso} = 0.033(1)$ Å$^2$ for all atoms. The occupancy for each atom was fixed at 1.

**Table 3. Results of Rietveld structure refinements for SrPd$_3$P and BaPd$_3$P at RT. Data of BaPd$_3$P are written in brackets.**

| atoms | Wyckoff position | x | y | z |
|---|---|---|---|---|
| Sr1 [Ba1] | 4a | 0 | 0 | 0.0118 [0.0089] |
| Sr2 [Ba2] | 4a | 0 | 1/2 | -0.0102(10) [-0.0215(6)] |
| Sr3 [Ba3] | 4a | 1/2 | 0 | -0.0255(5) [-0.0272(4)] |
| Sr4 [Ba4] | 4a | 1/2 | 1/2 | -0.0044(10) [-0.0047(5)] |
| Pd1 | 16c | 0.2474(5) [0.2462(4)] | 0.2489(3) [0.2484(7)] | 0.0254(6) [0.0225(4)] |
| Pd2 | 8b | 1/2 | 0.2722(5) [0.2374(7)] | 0.1147(7) [0.1101(6)] |
| Pd3 | 8b | 0 | 0.2871(5) [0.3264(5)] | 0.1191(7) [0.1177(5)] |
| Pd4 | 8b | 0 | 0.1772(5) [0.2218(6)] | 0.3689(7) [0.3625(6)] |
| Pd5 | 8b | 1/2 | 0.2536(6) [0.2280(6)] | 0.3727(6) [0.3657(6)] |
| P | 16c | 0.2453(10) [0.2478(13)] | 0.2511(10) [0.2585(12)] | 0.1560(7) [0.1550(4)] |

$I4_1md$ (tetragonal no. 109): For SrPd$_3$P, $a = 8.8064(2)$ Å, $c = 19.2228(6)$ Å, $V = 1491(1)$ Å$^3$ ($Z = 16$), and $R_{wp} = 8.51\%$, $R_e = 6.14\%$, $S = 1.39$. For BaPd$_3$P, $a = 8.9298(2)$ Å, $c = 19.5646(7)$ Å, $V = 1560(1)$ Å$^3$ ($Z = 16$), and $R_{wp} = 7.92\%$, $R_e = 5.32\%$, $S = 1.49$. $U_{iso}$ of all atoms were fixed to be 0.033(1) Å$^2$ and 0.04(1) Å$^2$ for SrPd$_3$P and BaPd$_3$P, respectively. The occupancy for each atom was fixed at 1.

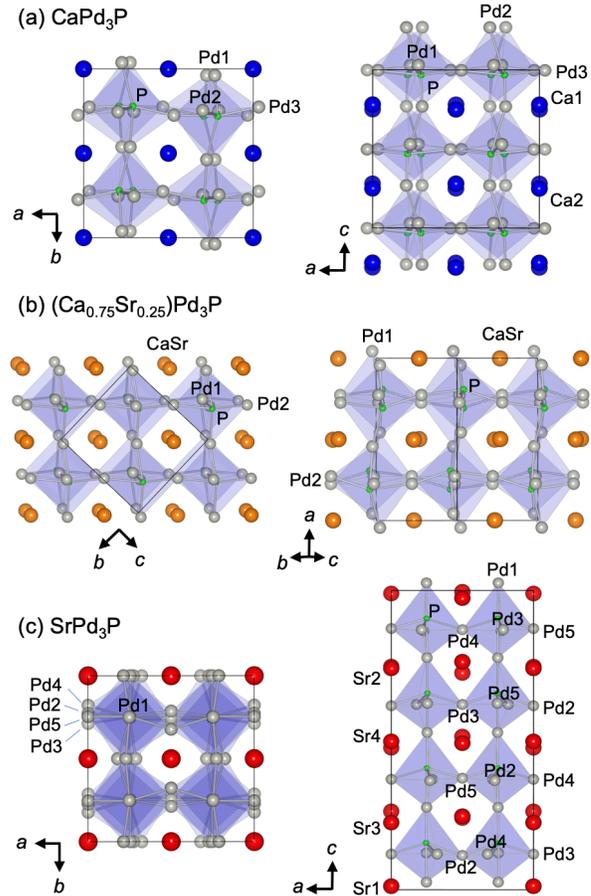

Figure 5. Schematic refined crystal structures of (a) NCS-orthorhombic CaPd$_3$P (150 K), (b) CS-orthorhombic (Ca$_{0.75}$Sr$_{0.25}$)Pd$_3$P, and (c) NCS-tetragonal SrPd$_3$P produced using the VESTA software.[7] Solid lines indicate the unit cells. Pd$_6$P octahedrons are shaded.

## Superconductivity in (Ca$_{1-x}$Sr$_x$)Pd$_3$P (0 ≤ $x$ ≤ 0.20)

Resistivity and magnetization data are displayed separately in the two composition regions of 0 ≤ $x$ ≤ 0.2 and 0.3 ≤ $x$ ≤ 1.0. Figure 6 shows the temperature ($T$) dependence of resistivity ($\rho$) and $4\pi M/H$ for representative $x$ (= 0, 0.1, 0.15, 0.17 and 0.2) in 0 ≤ $x$ ≤ 0.2. Resistivity anomalies with hysteretic behaviors were observed in the resistivity data for x = 0, 0.1, 0.15, and 0.17. The temperatures at which the resistivity changed abruptly in the cooling and warming processes were defined as $T_{sc}$ and $T_{sw}$, respectively, as indicated for $x$ = 0. The hysteresis loop closed above 320 K for CaPd$_3$P, which is the reason for the presence of the two orthorhombic phases at RT. The resistivity anomaly with the characteristic of a first-order transition can be attributed to the structural phase transition detected by the XRD measurements. $T_{sc}$ and $T_{sw}$ were determined to be 269 and 283 K, respectively. Then, $T_{sc}$ and $T_{sw}$ decreased with increasing x, and the anomaly and hysteresis disappeared for $x$ = 0.2.

Below the structural transition temperature, namely in the NCS-orthorhombic phase, the resistivity decreased more steeply with temperature. Consequently, the normal state resistivity of CaPd$_3$P decreased to 3.16 μΩcm (at 5 K), which was relatively small as a polycrystalline sample. A residual resistivity ratio (*RRR*) defined by $\rho(300K)/\rho(5K)$ was as large as ~80.2 for CaPd$_3$P. In contrast, the resistivity was less sensitive to temperature in the CS-orthorhombic phase e.g. *RRR* was only 1.6 for $x$ = 0.2.

The magnetization data clearly exhibited the evolution of superconductivity in (Ca$_{1-x}$Sr$_x$)Pd$_3$P. Superconducting transitions were more or less observed for all the samples in Figure 5. However, the transitions were broad and the superconducting shielding (ZFC) volume fraction at 2 K ($V_s$) were as small as ~7% and 35% for $x$ = 0 and 0.1, respectively. Transitions were still broad for $x$ = 0.15 though $V_s$ increased to ~100%. We believe that only the CS-orthorhombic phase is responsible for the superconductivity at ~3.5 K. The broad superconducting transitions at 0 ≤ $x$ ≤ 0.15 is probably caused by the CS-orthorhombic phase partially remaining in the samples as a result of the imperfect structural phase transition. Sharp transitions in both resistivity and magnetization began to appear at $x$ = 0.17. $T_c$ determined by resistivity and magnetization for $x$ = 0.17 as indicated in Figure 5 were 3.51 K and 3.32 K, respectively.

## Superconductivity in (Ca$_{1-x}$Sr$_x$)Pd$_3$P (0.3 ≤ $x$ ≤ 1.0) and BaPd$_3$P

Figure 7 shows the temperature dependence of resistivity and $4\pi M/H$ for representative $x$ (= 0.3, 0.55, 0.6, 0.8 and 1.0) in 0.3 ≤ $x$ ≤ 1.0 and BaPd$_3$P. Resistivity data were measured down to 20 mK for (Sr$_{0.75}$Ca$_{0.25}$)Pd$_3$P, SrPd$_3$P, and BaPd$_3$P. No anomalies were detected in the normal state resistivity; therefore, structural phase transitions appeared to be absent for the samples in Figure 7. Abrupt superconducting transitions were observed at ~3.5 K for samples (from $x$ = 0.17) up to $x$ = 0.55. $T_c$ determined by magnetization data were 3.32, 3.32, 3.33, 3.37, 3.44, and 3.47 K for $x$ = 0.17, 0.2, 0.3, 0.4, 0.5, and 0.55, respectively. Thus, $T_c$ slightly enhanced as $x$ increased.

In the NCS-tetragonal phase samples (0.6 ≤ $x$ ≤ 1.0, 0 ≤ $y$ ≤ 1.0), superconductivity suppressed drastically. The sample with $x$ = 0.6 exhibited only a trace of superconductivity at 3.5 K, most probably due to unavoidable compositional fluctuations in solid solution polycrystalline samples. $T_c$ decreased significantly down to 0.32 and 57 mK for $x$ = 0.75 and 1.0, respectively. Moreover, no superconductivity was observed above 20 mK in BaPd$_3$P. Thus, the NCS-tetragonal as well as the NCS-orthorhombic structures were unfavorable for superconductivity in this material. Note that *RRR*s of the endmembers SrPd$_3$P and BaPd$_3$P are large (10.6 and 11.9, respectively). Therefore, the disorder of lattice caused by the mixing of different cations may have resulted in the small *RRR* values (~2) of the solid solutions.

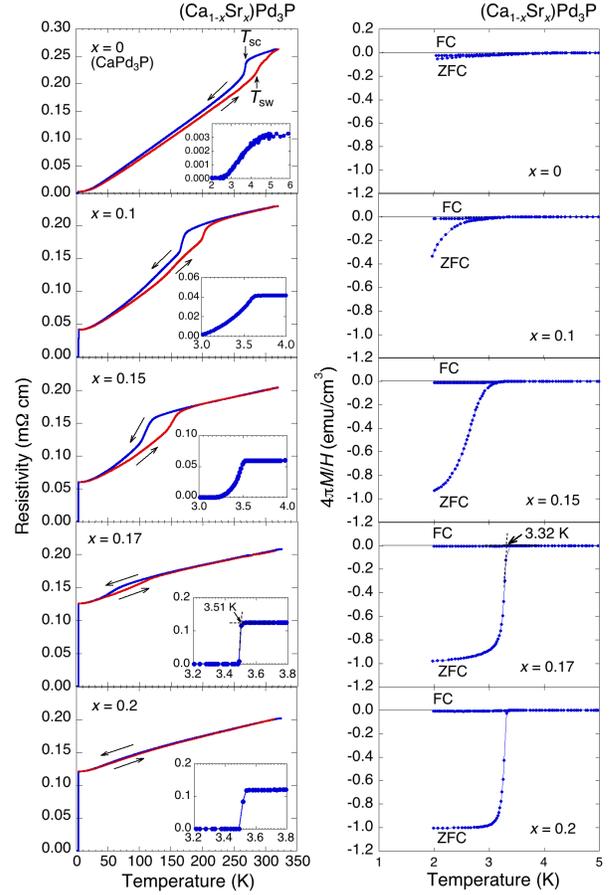

Figure 6. Temperature dependence of resistivity (left panel) and $4\pi M/H$ (right panel) for (Ca$_{1-x}$Sr$_x$)Pd$_3$P in the composition range 0 ≤ $x$ ≤ 0.20. The resistivity measurements with decreasing and increasing temperature are indicated by blue and red curves, respectively. Insets show increments in resistivity near superconducting transitions.

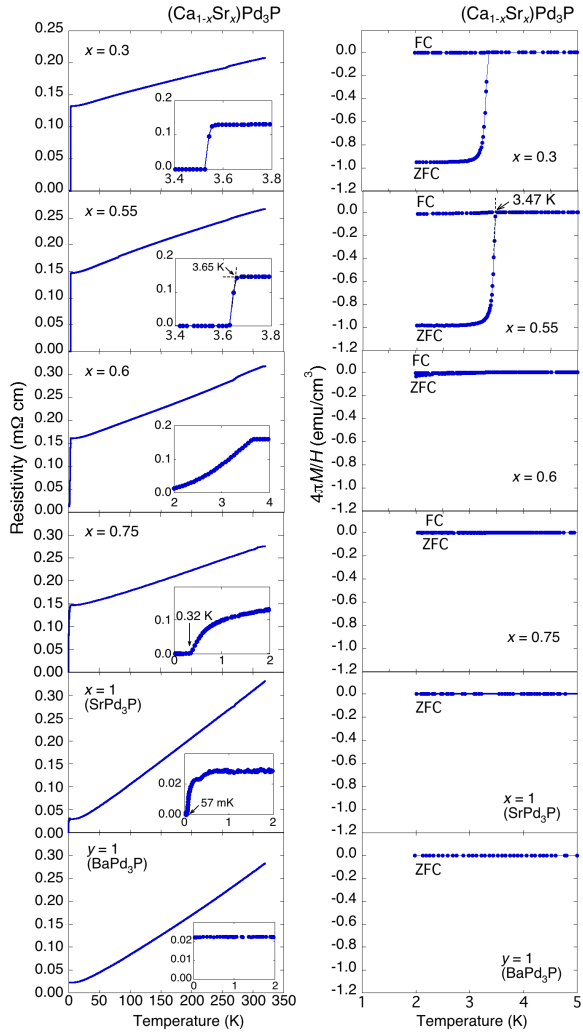

Figure 7. Temperature dependence of resistivity (left panel) and $4\pi M/H$ (right panel) for $(Ca_{1-x}Sr_x)Pd_3P$ in the composition range $0.3 \leq x \leq 1.0$ and $BaPd_3P$. Resistivities of $(Sr_{0.75}Ca_{0.25})Pd_3P$, $SrPd_3P$, and $BaPd_3P$ were measured down to 20 mK. Insets show resistivity in low-temperature regions.

**Phase diagram**

Schematic phase diagram of $(Ca_{1-x}Sr_x)Pd_3P$ drawn based on the experimental results is shown in Figure 8. There were two critical compositions regarding the structure and superconductivity. The endmember $CaPd_3P$ had the structural phase transition from CS- to NCS-orthorhombic system near RT. As $x$ increases, the transition temperature $T_s$ (= $T_{sc}/2 + T_{sw}/2$) decreases toward $x = 0.17$–$0.2$. The CS-orthorhombic phase was stable from $x = 0.17$–$0.2$ to $x = 0.55$. The structure then changed to the NCS-tetragonal system for further substitution ($0.6 \leq x < 1.0$ and $0 \leq y < 1.0$). Thus, the crystal structure (inversion symmetry) was precisely controlled by the substitution using electronically similar elements.

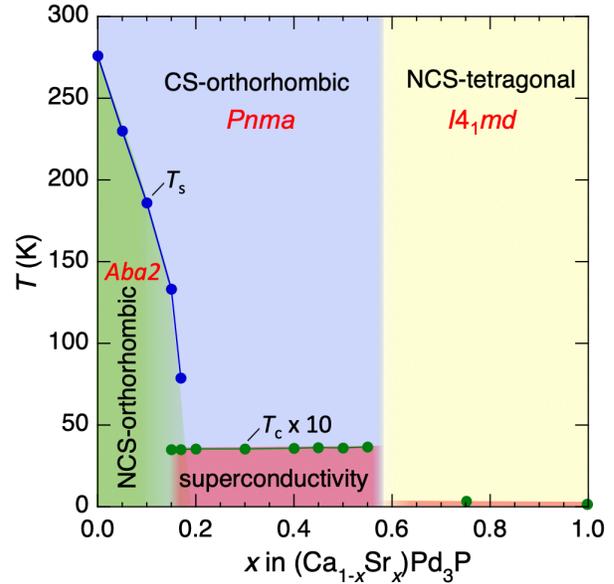

Figure 8. Schematic phase diagram of $(Ca_{1-x}Sr_x)Pd_3P$ drawn based on the powder XRD, resistivity, and magnetization measurements. $T_c$ in $0 \leq x \leq 0.6$ were plotted for samples with $-4\pi M/H > 0.5$ (ZFC) at 2 K.

Subedi et al. calculated the difference in total energies between the antipolar (CS) and polar (NCS) structures of $M'Pt_3P$ and $M'Pt_3Si$ ($M'$ = Sr, Ca, La).[18] They found that the energy differences in phosphides, especially $CaPt_3P$, are small, and indicated that the structure can be tuned from CS to NCS through appropriate synthesis conditions or partial replacement of elements. They showed that such structural control opens new perspectives toward the understanding of unconventional superconductivity. We realized their proposal in the $CaPd_3P$ system.

Hu et al. investigated the effect of Pd substitution for $SrPt_3P$.[19] They synthesized $Sr(Pt_{1-z}Pd_z)_3P$ for $0 \leq z \leq 0.4$ and found that $T_c$ decreases monotonously as $z$ increases. In addition, they found that the electron correlation is enhanced while the electron-phonon coupling and spin-orbit coupling are suppressed by Pd substitution. However, samples with $z > 0.4$ were not synthesized due to the substitution limit of Pd. If samples can be synthesized by optimizing synthesis conditions, there must be a phase transition from CS to NSC structure in $0.4 < z < 1.0$, which can be a suitable platform to study the effect of structure on superconductivity.

Bulk superconductivity at ~3.5 K began to occur in the CS-orthorhombic phase at $x = $ ~0.17. Note that the sample of $x = 0.17$ exhibited the normal state resistivity anomaly. However, the anomalies are smaller than those of other samples, suggesting that only a small portion of the sample undergoes a structural phase transition, and most of the sample retains the CS-orthorhombic phase. $T_c$ monotonically and slightly increased with the composition $x$ and reached a maximum value of 3.65 K (determined by resistivity data for $x = 0.55$) just before the phase boundary. Then, superconductivity was significantly suppressed in the NCS-tetragonal phase. $(Ca_{0.25}Sr_{0.75})Pd_3P$ and $SrPd_3P$ showed zero resistivity at as low as 0.32 K and 57 mK, respectively. The lack of inversion symmetry appears to be fatal to superconductivity in $(Ca_{1-x}Sr_x)Pd_3P$. This material may offer a unique opportunity to study the effect of inversion symmetry on superconductivity which is currently attracting considerable attention in the field of superconductivity.[20,21]

It is currently unclear why superconductivity varies significantly depending on the crystal structure and how inversion symmetry is related to superconductivity in this material. Detailed physical property measurements and theoretical work such as band structure calculations are necessary to obtain a better understanding regarding this. Finally, we stress that there is room to explore new antiperovskite Pd phosphides in combination with other cations such as alkali metals and rare earth elements. In addition, the previously reported Zr(P,Se)$_2$ (PbClF-type) superconductor was also observed in the solid solution between ZrP$_2$ (PbCl$_2$-type) and ZrSe$_2$ (CdI$_2$-type).[22] Therefore, searching for superconductivity in such a hidden phase can also lead to the discovery of new materials/superconductors.

## CONCLUSIONS

In our attempt to discover new antiperovskite phosphide materials, we succeeded in synthesizing new $M$Pd$_3$P compounds and discovered a superconducting phase in the solid solutions between CaPd$_3$P and SrPd$_3$P. Three perovskite-related structures were identified in (Ca$_{1-x}$Sr$_x$)Pd$_3$P and (Sr$_{1-y}$Ba$_y$)Pd$_3$P, and superconductivity was found to depend strongly on the structures, possibly because of the presence of inversion symmetry. After the suppression of the NCS-orthorhombic phase by the Sr substitution, superconductivity was induced in the CS-orthorhombic phase (0.17 ≤ $x$ ≤ 0.55) at ~3.5 K. For further substitution (0.6 ≤ $x$ ≤ 1.0 and 0 ≤ $y$ ≤ 1.0), the NCS-tetragonal phase was stable, and the superconductivity was significantly suppressed. In order to understand the effect of the crystal structure variation on superconductivity, further experiments with theoretical support are currently in progress.

## ACKNOWLEDGMENTS

This work was supported by the JSPS KAKENHI (No. JP19K04481 and JP19H05823) and TIA collaborative research program KAKEHASHI "Tsukuba-Kashiwa-Hongo Superconductivity Kakehashi Project".